\documentclass[aps,pra,twocolumn,showpacs,amsmath,amssymb,preprintnumbers,superscriptaddress,10pt]{revtex4-1}

\usepackage{graphicx}
\usepackage{hyphenat}
\usepackage{url}
\usepackage[bookmarks=false]{hyperref}
\usepackage[capitalise]{cleveref}
\crefname{section}{Sec.}{Secs.}
\Crefname{section}{Section}{Sections}
\usepackage[retain-explicit-plus]{siunitx}
\usepackage[titletoc]{appendix}

\begin{document}

\title{Repeated radiation damage and thermal annealing of avalanche photodiodes}

\author{Ian~DSouza}
\affiliation{Institute for Quantum Computing, University of Waterloo, Waterloo, ON, N2L~3G1 Canada}
\affiliation{Department of Physics and Astronomy, University of Waterloo, Waterloo, ON, N2L~3G1 Canada}

\author{Jean-Philippe~Bourgoin}
\affiliation{Present address: Aegis Quantum, Waterloo, ON, Canada}
\affiliation{Institute for Quantum Computing, University of Waterloo, Waterloo, ON, N2L~3G1 Canada}
\affiliation{Department of Physics and Astronomy, University of Waterloo, Waterloo, ON, N2L~3G1 Canada}

\author{Brendon~L.~Higgins}
\email{brendon.higgins@uwaterloo.ca}
\affiliation{Institute for Quantum Computing, University of Waterloo, Waterloo, ON, N2L~3G1 Canada}
\affiliation{Department of Physics and Astronomy, University of Waterloo, Waterloo, ON, N2L~3G1 Canada}

\author{Jin~Gyu~Lim}
\affiliation{Institute for Quantum Computing, University of Waterloo, Waterloo, ON, N2L~3G1 Canada}
\affiliation{\mbox{Department of Electrical and Computer Engineering, University of Waterloo, Waterloo, ON, N2L~3G1 Canada}}

\author{Ramy~Tannous}
\affiliation{Institute for Quantum Computing, University of Waterloo, Waterloo, ON, N2L~3G1 Canada}
\affiliation{Department of Physics and Astronomy, University of Waterloo, Waterloo, ON, N2L~3G1 Canada}

\author{Sascha~Agne}
\affiliation{Institute for Quantum Computing, University of Waterloo, Waterloo, ON, N2L~3G1 Canada}
\affiliation{Department of Physics and Astronomy, University of Waterloo, Waterloo, ON, N2L~3G1 Canada}

\author{Brian~Moffat}
\affiliation{Institute for Quantum Computing, University of Waterloo, Waterloo, ON, N2L~3G1 Canada}

\author{Vadim~Makarov}
\affiliation{Russian Quantum Center, Skolkovo, Moscow 121205, Russia}
\affiliation{\mbox{Shanghai Branch, National Laboratory for Physical Sciences at Microscale and CAS Center for Excellence in} \mbox{Quantum Information, University of Science and Technology of China, Shanghai 201315, People's Republic of China}}
\affiliation{NTI Center for Quantum Communications, National University of Science and Technology MISiS, Moscow 119049, Russia}
\affiliation{Department of Physics and Astronomy, University of Waterloo, Waterloo, ON, N2L~3G1 Canada}

\author{Thomas~Jennewein}
\affiliation{Institute for Quantum Computing, University of Waterloo, Waterloo, ON, N2L~3G1 Canada}
\affiliation{Department of Physics and Astronomy, University of Waterloo, Waterloo, ON, N2L~3G1 Canada}

\begin{abstract}
Avalanche photodiodes (APDs) are well-suited for single-photon detection on quantum communication satellites as they are a mature technology with high detection efficiency without requiring cryogenic cooling. They are, however, prone to significantly increased thermal noise caused by in-orbit radiation damage. Previous work demonstrated that a one-time application of thermal annealing reduces radiation-damage-induced APD thermal noise. Here we examine the effect of cyclical proton irradiation and thermal annealing. We use an accelerated testing environment which emulates a realistic two-year operating profile of a satellite in low-Earth-orbit. We show that repeated thermal annealing is effective at maintaining thermal noise of silicon APDs within a range suitable for quantum key distribution throughout the nominal mission life, and beyond. We examine two strategies---annealing at a fixed period of time, and annealing only when the thermal noise exceeds a pre-defined limit. We find both strategies exhibit similar thermal noise at end-of-life, with a slight overall advantage to annealing conditionally. We also observe that afterpulsing probability of the detector increases with cumulative proton irradiation. This knowledge helps guide design and tasking decisions for future space-borne quantum communication applications.
\end{abstract}

\maketitle

\section{Introduction}
\label{sec:introduction}

Distributing secret keys to users of a communication network enables them to exchange messages securely using cryptographic protocols. To do so, conventional cryptosystems utilize algorithms that provide data security under assumptions of computational complexity. Quantum computers, however, are in principle capable of efficiently solving problems that are foundational to some of these algorithms (by using, e.g., Shor's factoring algorithm~\cite{shorAlgorithm}). ``Post-quantum'' encryption protocols such as the McEliece protocol~\cite{mcEliece} attempt to be immune to attacks by a quantum computer, but their robustness to cryptanalysis is not well established---in some cases being prone to even classical attacks~\cite{attackOnMcEliece}.
Quantum key distribution (QKD) protocols~\cite{bennett1984, BBM92, scarani2009qkd} make use of public quantum channels to securely distribute symmetric keys by exchange of quantum bits. Fundamentally, security is obtained through the no-cloning theorem: any measurement of a quantum bit by an eavesdropper risks changing the state of that bit, through which the eavesdropper's presence is revealed.

QKD is typically implemented using photons which travel through optical fibers or over atmospheric free space. Both experience high losses as transmission distances increase, with the dominant losses scaling exponentially. This makes it impossible to distinguish transmitted signals above intrinsic photon detector noise beyond distances of a few hundred kilometers. In contrast, orbiting satellites have the advantage that optical transmission through the vacuum of space is dominated by beam divergence, where loss scales quadratically. Thus, quantum satellites, acting as nodes, can help extend the range of QKD. Moreover, placing photon detectors on such a satellite (as opposed to on the ground) has advantages of relaxing its pointing, computational, and memory requirements. The same detector apparatus can additionally be used in a wider range of scientific experiments~\cite{bourgoin2013comprehensive}. Such an approach has been trialled with the Micius satellite~\cite{liao2017} and is being pursued by the QEYSSat satellite~\cite{jennewein2014, qeyssat2020}.

For single-photon detection, silicon avalanche photodiodes (Si APDs) have the advantage of low dark count rates, high maximum count rates, and high photon detection efficiencies when coupled to the near-infrared wavelengths around \SI{800}{\nm} that are a good compromise between atmospheric transmittance and diffraction~\cite{bourgoin2013comprehensive}. In addition, Si APDs do not require cryogenic cooling, as competing detector technologies do, greatly simplifying support systems that would be required to operate on a space-borne platform. However, proton irradiation due to solar particle events~\cite{chancellor2014space} can induce displacement damage~\cite{displacementDamage} that causes defects in the Si APD substrate placed in low-Earth orbit (LEO). This leads to noise in the form of dark counts within these detectors, thereby decreasing their signal-to-noise ratios throughout the operational lifetime of a spacecraft payload. Thermal annealing---where a detector's active area is heated to remove crystalline defects---has been proposed as a technique to mitigate such radiation damage and decrease dark count rates. This has shown to be effective in single-shot experiments~\cite{ThesisElena} when coupled with operating the detectors at cold temperatures, e.g., \SI{-80}{\celsius}. For a satellite platform, such cold operation can be achieved by thermoelectric cooling supported by passive thermal radiation to deep space. At the same time, annealing temperatures can be achieved by thermoelectric heating supported by resistive electric heating and orbital manoeuvres.

In this work, we examine the effects of repeated thermal annealing on detector parameters over the emulated life of a satellite payload detector apparatus. We irradiate Si APDs with protons over several stages and measure the detector performance at each stage, accumulating dosage equivalent to at least two years in LEO. We demonstrate the efficacy of repeated thermal annealing to reduce dark count rates to within operational range~\footnote{Many parameters, which vary according to the orbital environment and mission requirements, influence the operational performance of a QKD system. The scenario we consider follows that of Ref.~\cite{bourgoin2013comprehensive}.}.

To determine suitable criteria for engaging annealing, we study two strategies. The first is to anneal the APDs periodically, regardless of current performance. The second is to only anneal when the dark count rate exceeds a pre-determined threshold value (here \SI{2}{\kilo cps}, kilo counts per second). We thus irradiate two identical sets of detectors using these two approaches and compare the resulting detector performance. Our results show that although both approaches achieve similar outcomes, the case where the detectors are annealed only when the dark count rate exceeds a threshold value generally achieves slightly lower post-annealed dark count rates at the end of the two-year life of the satellite.

This paper is organized as follows. In \cref{sec:setup}, we describe the experimental setup to irradiate the detectors and monitor detector parameters. We present our test methodology in \cref{sec:methodology} to examine the effects of irradiation and annealing on detector parameters. In \cref{sec:results}, we present our results and inferences of the irradiation and annealing experiments. Concluding remarks are given in \cref{sec:conclusion}.

\section{Experimental setup}
\label{sec:setup}

\begin{figure}
	\includegraphics[width=\linewidth]{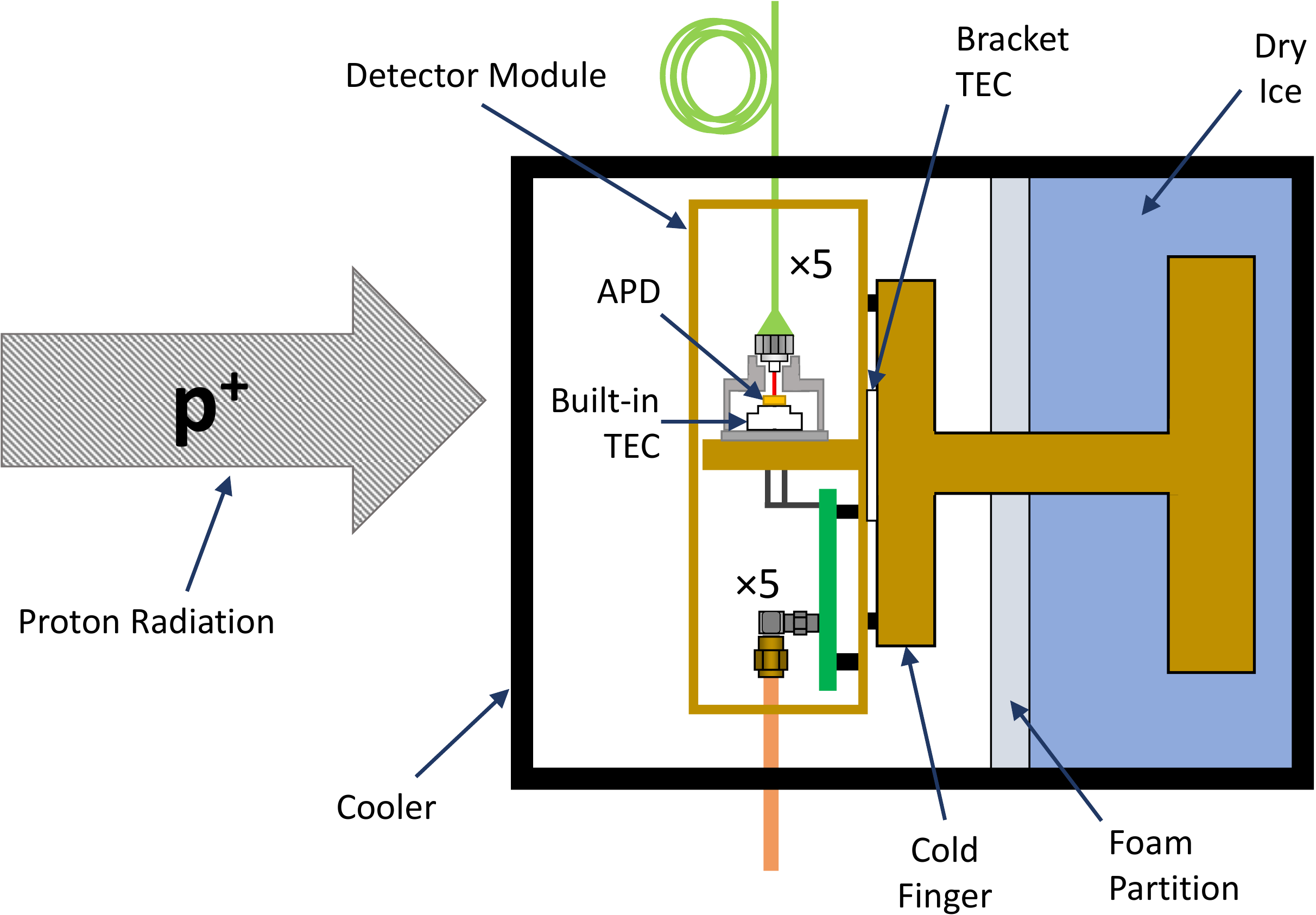}
	\caption[Schematic of detector module in cooler]{Cross-section schematic of the detector module (DM) inside a passive cooler box. A cold finger thermally interfaces dry ice with the DM. The DM and cold finger assembly is mounted on its side such that protons, entering the cooler through the side wall, are minimally disturbed by various mechanical structures prior to incidence on the detectors (remaining structures are accounted for in our fluence model). Drawing is not to scale; smaller parts are exaggerated to show detail.
	} 
	\label{fig.dmPhoto}
\end{figure}

Five silicon APDs manufactured by Excelitas (the packaged APD unit normally found inside the commercial detector module SPCM-AQRH-12-FC) are housed in a custom enclosure together with supporting electronics designed for operating in space. This unit is called the detector module (DM). The APDs are mounted on a bracket which is thermally isolated from the enclosure and connected to a heat sink. Each APD's active area is temperature stabilized using a thermo-electric cooler (TEC) that is housed inside the APD package. The detector TEC is thermally interfaced to the bracket and driven by closed-loop control based on thermistor readout. Another TEC mounted between the bracket and the heat sink serves to stabilize the temperature of the bracket and helps achieve a wider range of APD operating temperatures.

Nominally the heat sink would be mounted to a radiator pointed towards deep space to maintain a low temperature. We emulate this for our tests---the heat sink is maintained at a temperature of about \SI{-78.5}{\celsius} by thermally interfacing it to one end of a copper cold finger whose other end is embedded in dry ice. This whole setup is housed inside a passive cooler box (\cref{fig.dmPhoto}) to thermally isolate it from the ambient environment. The DM enclosure and dry ice are physically separated by an insulating foam partition. The detectors are typically operated around \SI{-80}{\celsius}, but for annealing they are heated by operating the bracket and APD TECs in reverse, to \SI{-40}{\celsius} and \SI{+80}{\celsius}, respectively.

\begin{figure}
	\includegraphics[width=\linewidth]{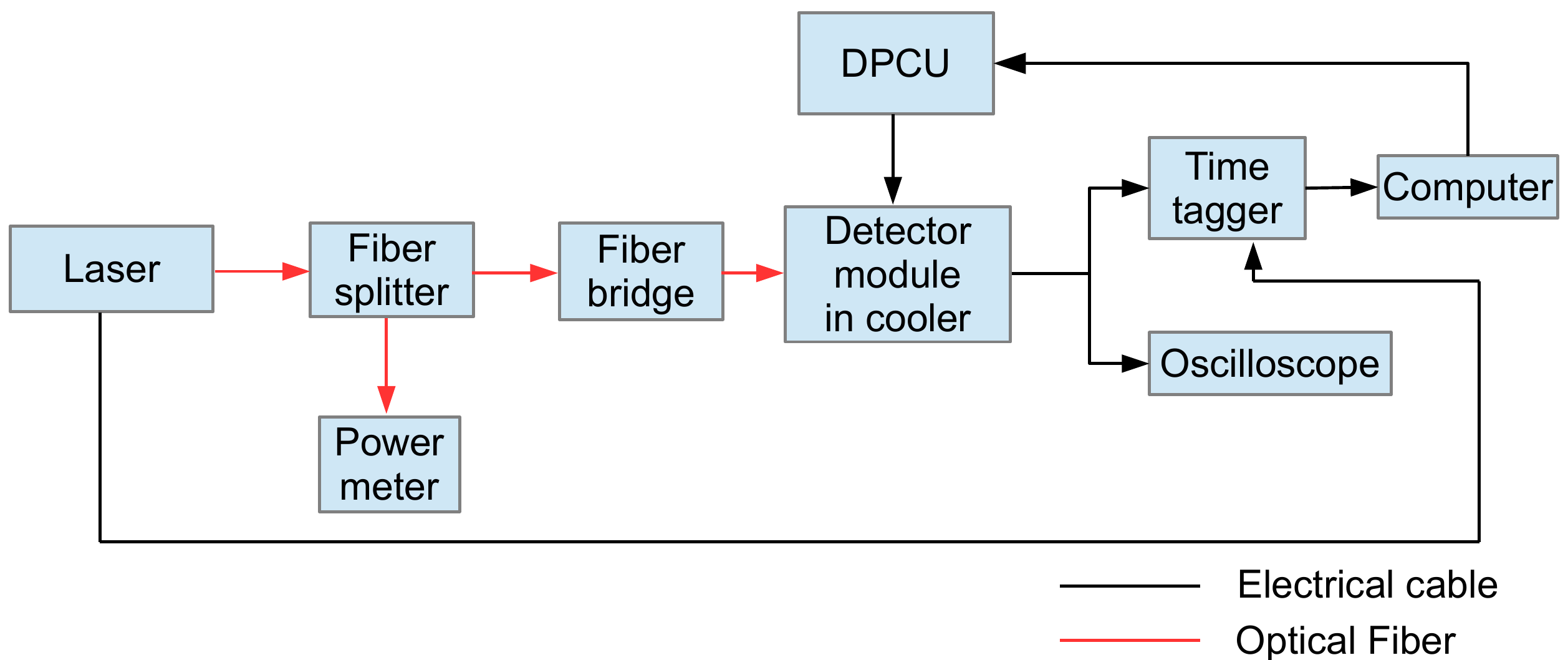}
	\caption[Schematic of experimental setup]{Schematic of the detector characterization setup. Si APDs, housed in a detector module, are placed inside a passive cooler box filled with dry ice acting as a heat sink. The detector power and control unit (DPCU) maintains detector bias voltages and controls thermal conditions via detector and bracket TECs. Fibers guide optical pulses from a laser to the detectors and to an optical power meter for reference measurement via a fiber beam splitter. A fiber bridge coupling into a multi-core fiber is used to split the laser light between each of the detectors. The output from the detectors is either measured on an oscilloscope or time-tagged and recorded on a computer.}
	\label{fig.schematicSetup}
\end{figure}

A fiber-coupled pulsed laser emitting \SI{785}{\nm} wavelength light is used to characterize APD performance (\cref{fig.schematicSetup}). A fiber beam splitter placed in the optical path diverts a small portion of the laser power to a pre-calibrated optical power meter that monitors the incident laser power, while the remaining signal is coupled via a free-space fiber bridge into a 1-to-19 fiber bundle. Five outputs of the fiber bundle which have relatively matched optical powers are used such that attenuated laser power is applied uniformly to each APD. As our primary objective is to detect the relative change of efficiency due to irradiation, we need only ensure the same fiber configuration is used throughout the measurement.

The APDs are run in Geiger mode, biased \SI{20}{\volt} above their respective breakdown voltages, and use a passive quenching circuit~\cite{ThesisElena, passiveQuenching, lim2017, kim2011, GeigerMode}. Electrical output pulses of each APD are measured either on an oscilloscope or using a time tagging unit. The oscilloscope is used to measure breakdown voltage, output pulse amplitude and width, and recharge time. The time tagging unit measures the arrival time of each output pulse. Appropriate post-processing algorithms applied to the recorded time tags then allow evaluation of the detector dark count rate, photon detection efficiency, timing jitter, maximum count rate (saturation), and afterpulsing probability.

We consider the scenario of two years in LEO at \SI{600}{\kilo\metre} altitude, and assume \SI{10}{\mm} aluminum shielding from the spacecraft walls. Following Ref.~\cite{Elena'sRadiationPaper}, which used a commonly accepted silicon damage deposition model~\cite{Dale1994}, under these conditions the cumulative fluence incident on the detectors is equivalent to about \SI{4e9}{protons\per\cm^2} (\si{p\per\cm^2}) at \SI{100}{\mega\electronvolt} energies~\cite{ThesisJin, ThesisElena}. We expose DMs to various proton fluences at the Tri-University Meson Facility (TRIUMF), producing protons with energies of about \SI{105}{\mega\electronvolt}. In our apparatus, an incoming proton will deposit some of its energy in the surrounding material---namely, the cooler box, which would be absent on the satellite, and a metal fiber coupler---prior to reaching an APD. Consequently, a \SI{105}{\mega\electronvolt} incident proton reduces to \SI{92.8}{\mega\electronvolt} at the APD~\cite{ThesisJin}. The reduced velocity leads to increased travel time through the APD substrate, where the proton then deposits a greater amount of energy, and the corresponding displacement damage is 1.052 times greater than what would be obtained with an incident proton energy of \SI{100}{\mega\electronvolt} at the APD.

\section{Methodology}
\label{sec:methodology}

Three detector modules were constructed and brought to TRIUMF for testing conducted May 1--6, 2017. Two of these modules, DM1 and DM2, underwent multiple rounds of proton irradiation to emulate the gradual damage accumulation and cyclical usage that would be experienced in orbit. The third DM was not irradiated and acts as a control.

Proton fluences were applied to DM1 in regular increments of nominally \SI{6.67e8}{p/\cm^2}, corresponding to 4.2 months equivalent in LEO after adjusting for energy loss in materials surrounding our APDs. We thermally annealed DM1 for one hour between each irradiation. (For more on the annealing duration, see Appendix.) In contrast, thermal annealing of DM2 (also one hour) was only performed if the dark count rate exceeded \SI{2}{\kilo cps} in each of two or more detectors, to mitigate against outliers. The dark count rate threshold is chosen in view of QKD performance projected for our modelling of ground to space link---see Ref.~\cite{bourgoin2013comprehensive}. Taking this approach, we can compare the effects of the two annealing strategies, periodic and conditional. To allow more fine-grained observation of the dark count rate, proton fluences were applied to DM2 in irregular, initially smaller increments.

DM1 was irradiated to a total fluence of \SI{4e9}{p/\cm^2} (2.1 year equivalent in LEO) while DM2 was irradiated to a total fluence of \SI{2e10}{p/\cm^2} (10.5 year equivalent in LEO), with annealing performed at the nominal two-year mission end as a matter of course for comparison to DM1. We subjected DM2 to greater cumulative fluence---well beyond the nominal mission duration---and colder operating temperatures to explore potential late-life performance of the system. All three DMs were characterized before and after shipping to TRIUMF, while DM1 and DM2 were also characterized at TRIUMF before initial irradiation, after each irradiation (prior to annealing, if any), and again after each annealing (where applicable). Each characterization comprised measuring breakdown voltage, output pulse amplitude and full width at half maximum (FWHM), recharge time, dark count rate, photon detection efficiency, timing jitter, saturation, and afterpulsing probability, for each detector at \SI{-80}{\celsius}.

Characterization of each DM proceeded as follows: the breakdown voltage of each detector was measured by observing the presence of pulses on an oscilloscope. At \SI{20}{\volt} above the breakdown voltage, the detector's output pulse amplitude and FWHM, and recharge time were then observed by eye over the oscilloscope and recorded. Time-tags of output pulses were then recorded (for all detectors at once) with and without the pulsing laser active. Finally, time-tags were recorded while the input laser power was slowly swept through saturation.

Dark count rates and detection efficiencies of each detector were determined from counts rates observed in inactive- and active-laser data, respectively. Detector timing jitter was extracted from the active-laser data by producing a histogram of time differences between detection events and laser reference pulses. The overall system timing jitter was taken as the FWHM of a Gaussian curve fit of this distribution, and jitter contributions of the laser and time tagging unit were then subtracted (in quadrature).

The active-laser data consists of photon and dark count detections, and afterpulses caused by each of those. Because the afterpulsing time constant is greater than the laser pulse period, this results in a time-delta histogram with a (roughly Gaussian) photon detection peak on top of an approximately constant noise floor. By performing similar analysis of the inactive-laser data, we determined the combined dark-count and associated afterpulsing contribution to this noise floor. Subtracting this from the active-laser time-delta histogram, we are left with contributions from only incident photons, at a rate $R_\text{photons}$, and a residual noise floor due to their associated afterpulses. We define the afterpulsing probability as
\begin{equation} \label{eqn.afterpulsing}
P_A = \frac{R_\text{afterpulsing}}{R_\text{photons} + R_\text{afterpulsing}}\ ,
\notag
\end{equation}
where the afterpulsing rate $R_\text{afterpulsing}$ includes only afterpulses that are caused by photon detection events. Numerator and denominator are evaluated from suitable off-peak and on-peak locations (respectively) in the time-delta histograms. The above definition corresponds to the average number of afterpulses directly arising due to the avalanche corresponding to any arbitrarily chosen output pulse. Although $P_A$ is mathematically evaluated with respect to the photon rate here, it can be generalized to any output pulse (including, e.g., afterpulses of afterpulses) on the assumption that all avalanches in the detector substrate are statistically indistinguishable from each other.

\section{Results and discussion}
\label{sec:results}

As expected, no property of the control DM showed any significant change for the duration of the experiment. For DM1 and DM2, we find that breakdown voltage, output pulse amplitude and FWHM, recharge time, photon detection efficiency, timing jitter, and saturation characteristics are generally uncorrelated with cumulative proton radiation fluence and thermal annealing. This is consistent with the results of previous studies~\cite{Elena'sRadiationPaper, DautetParametersUncorellatedWithRadiation}.

\begin{figure}
  \includegraphics[width=\linewidth]{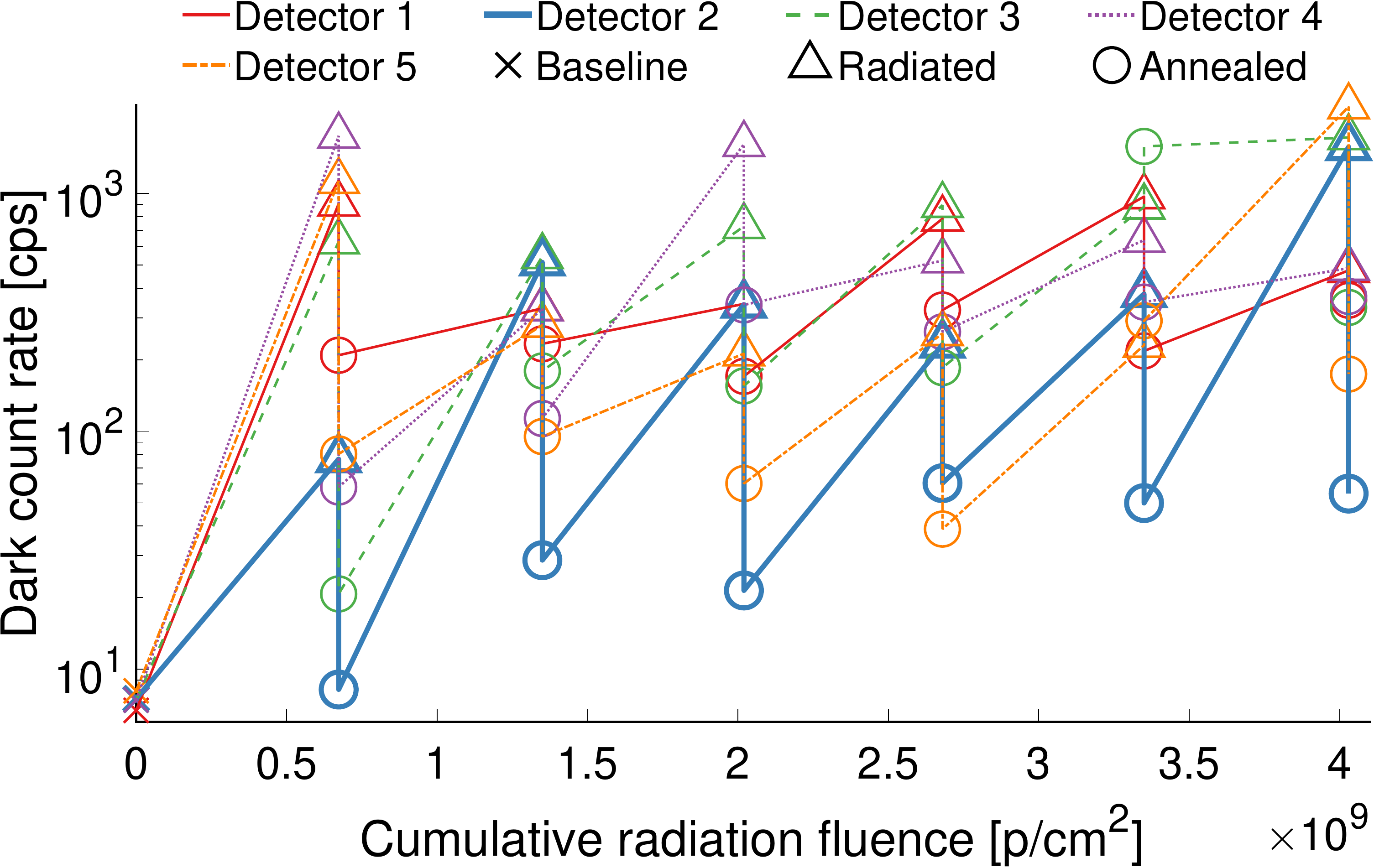}
  \caption[Dark counts of Radiation~1 across tests]{Dark count rates measured on DM1 after each irradiation and annealing for increasing cumulative proton radiation fluence. All characterizations are performed at a detector temperature of \SI{-80}{\celsius}. Following each incremental irradiation, DM1 is thermally annealed at \SI{+80}{\celsius} for one hour. In general, dark count rates are found to increase after each incremental irradiation and decrease after each annealing. Uncertainties, quantified by one standard deviation of Poissonian dark counts about the observed value, are smaller than the size of the marker symbols used in the graph.}
  \label{fig.Dark_test_rad_1}
\end{figure}

\begin{figure}
  \includegraphics[width=\linewidth]{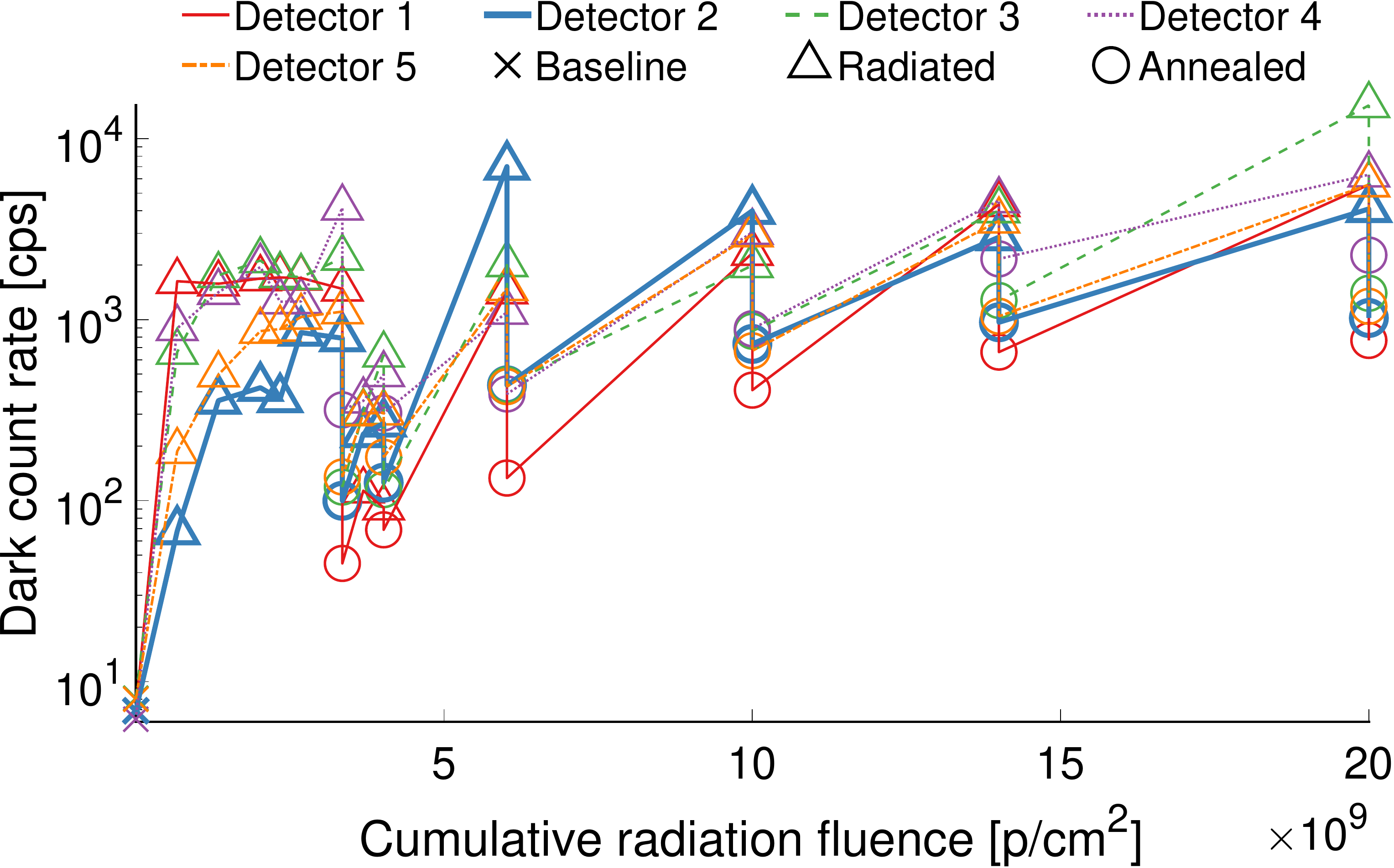}
  \caption[Dark counts of Radiation~2 across tests]{Dark count rates measured on DM2 after each irradiation and conditional annealing for increasing cumulative proton radiation fluence. All characterization is performed at a detector temperature of \SI{-80}{\celsius}. Thermal annealing at \SI{+80}{\celsius} for one hour is performed only if the measured dark count rate after irradiation exceeds \SI{2}{\kilo cps} in at least two detectors, and at \SI{4e9}{p/\cm^2} (nominal end-of-mission---see text). In general, dark count rates are found to increase after each incremental irradiation and decrease after each annealing. Uncertainties, quantified by one standard deviation of Poissonian dark counts about the observed value, are smaller than the size of the marker symbols used in the graph.}
  \label{fig.Dark_test_rad_2}
\end{figure}

\Cref{fig.Dark_test_rad_1} shows dark count rates measured after each incremental irradiation and fixed-period annealing of DM1. As expected, proton irradiation increases dark count rates, and we observe a decrease in dark count rates after each annealing, discounting two statistical anomalies at \SI{3.35e9}{p/cm^2} cumulative fluence. A general increase in post-annealed dark count rates with the cumulative fluence is apparent, indicating accumulated damage that annealing cannot repair. Nevertheless, this strategy maintains post-annealing dark count rates under \SI{500}{cps} throughout the 2.1-year-LEO-equivalent fluence for four of the five detectors. (The outlying detector experienced a post-annealing dark count rate increase, reaching about \SI{1.6}{\kilo cps} at \SI{3.35e9}{p/cm^2} cumulative fluence.) From prior analysis we may infer that these sub-\SI{500}{cps} dark count rates would lead to reduced QKD key rates by a few percent to a few tens of percent, relative to the nominal value (see Table~2 of Ref.~\cite{bourgoin2013comprehensive} and Table~2.6 of Ref.~\cite{ThesisJP}).

\Cref{fig.Dark_test_rad_2} shows dark count rates measured after each irradiation and conditional annealing of DM2. We again observe a general increase in dark count rates after each irradiation, although unlike for DM1, some instances of decreased dark count rate can be seen. (For example, three of the five detectors see a dark count rate decrease immediately after irradiation at \SI{2.35e9}{p/cm^2} cumulative fluence. Absent any indication of a causative mechanism, we attribute this to statistical fluctuations~\footnote{It is notable that six such anomalous results occur following a relatively small fluence increment of nominally \SI{3.33e8}{p/cm^2}, three occur following a \SI{6.67e8}{p/cm^2} fluence increment, and none are observed for higher fluence increments.}.)
Notably, the conditional annealing strategy saw annealing only once prior to the cumulative fluence equivalent to the two-year nominal mission life, at which point DM2 was annealed as a matter of course.
 
Like DM1, dark count rates decrease when annealing is applied. We find that thermal annealing is effective in maintaining dark count rates below \SI{500}{cps} for cumulative fluence up to \SI{6e9}{p/\cm^2}, 3.15-year-LEO-equivalent dosage. However, at the next cumulative fluence, \SI{1e10}{p/\cm^2} or 5.25-year-LEO-equivalent, the post-annealed dark count rate increases to over \SI{800}{cps} in two of the detectors.

We define the dark count reduction factor (DCRF) as the ratio of the dark count rate before annealing to that after annealing. Most DCRF values cluster between 2 and 5; the largest DCRF found is 32 on DM2. Annealing tends to have the largest DCRF when applied for the first time, when there is the least permanent radiation damage. After this the DCRF values are largely uncorrelated with cumulative fluence. The DCRF also did not show any correlation with the dark count rates measured before the annealing process. 

\begin{figure}
  \includegraphics[width=\linewidth]{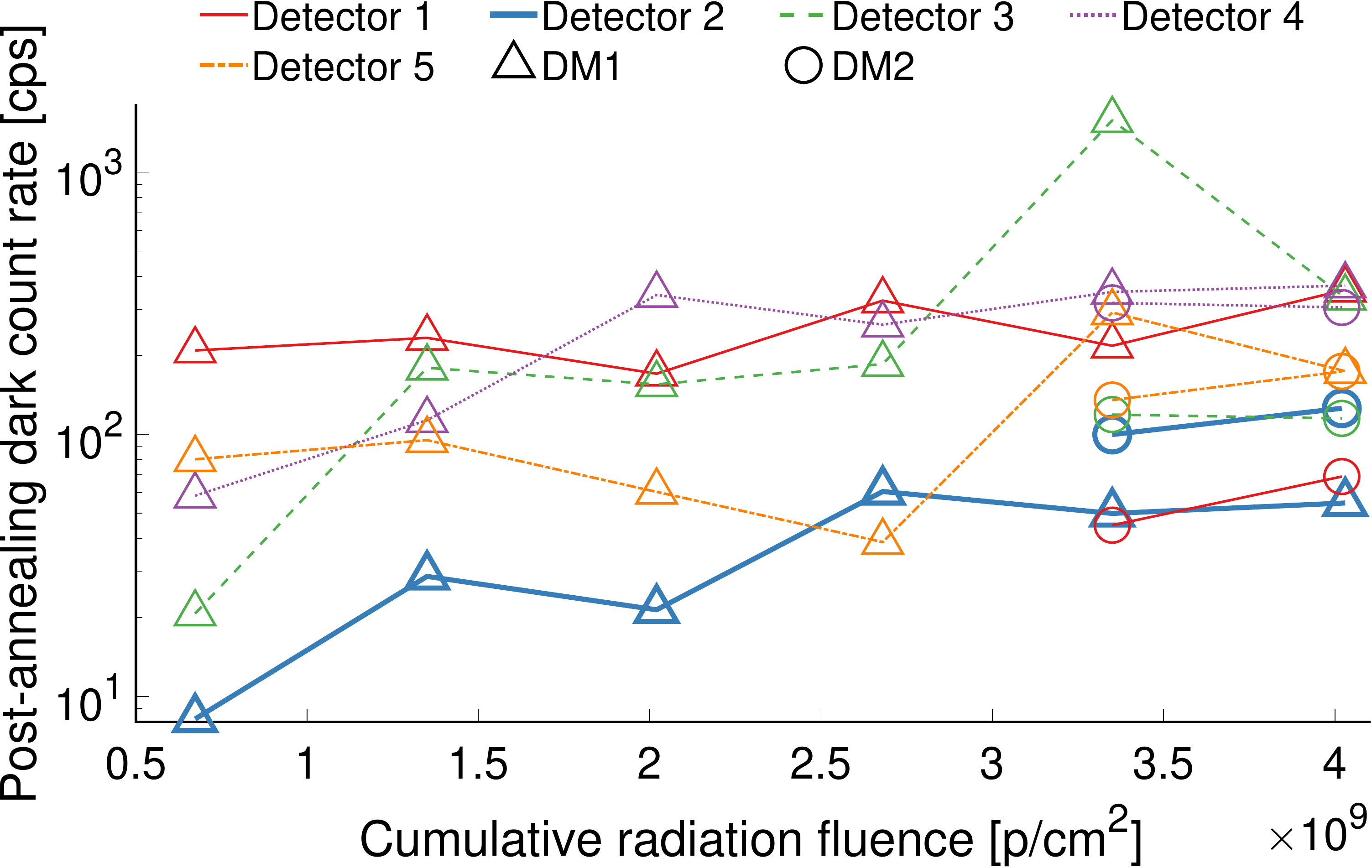}
  \caption[Comparison of the annealing methods]{Post-annealing dark count rates. DM1 is thermally annealed after each incremental irradiation, whereas DM2 is annealed only if the post-irradiated dark count rate exceeds \SI{2}{\kilo cps} on at least two detectors. Comparing the post-annealed dark count rates at \SI{3.35e9}{p/cm^2} and \SI{4e9}{p/cm^2} cumulative fluences, DM2 evidently achieves lower values across most detectors compared to DM1.}
  \label{fig.Post_annealed_dark_comparison}
\end{figure}

\Cref{fig.Post_annealed_dark_comparison} shows the post-annealed dark count rates for both DMs irradiated up to a cumulative fluence of \SI{4e9}{p/\centi\meter^2}. The relative performance of the two annealing strategies can be assessed from observations at the two cumulative fluences at which both DMs are annealed. The results indicate that DM2 has slightly lower post-annealed dark count rates across most detectors, despite both DMs being subjected to the same cumulative fluence. Due to the limited samples, it is possible that this originates from statistical variation; alternately, if real, then DM1 experienced worse performance as a consequence of being annealed more times. This suggests that it may be wise to anneal with some conservatism. Given that conditional annealing, with a suitably chosen threshold, will at least achieve the same overall dark count rates that are maintained using periodic annealing, it thus follows that conditional annealing should be preferred.

Irradiating DM2 into late-life condition, with a cumulative fluence equivalent to 10.5 years in LEO (\SI{2e10}{p/\cm^2}), we observe a mean detector dark count rate following annealing of \SI{1.3}{\kilo cps}. It is remarkable that the dark count rate can be maintained below the annealing threshold after such significant exposure. Despite worsening performance, prior calculations~\cite{bourgoin2013comprehensive, ThesisJP} indicate that DM2's dark count rates would still be low enough to perform QKD even at this well-past-nominal late-life stage.

\begin{figure}
  \includegraphics[width=\linewidth]{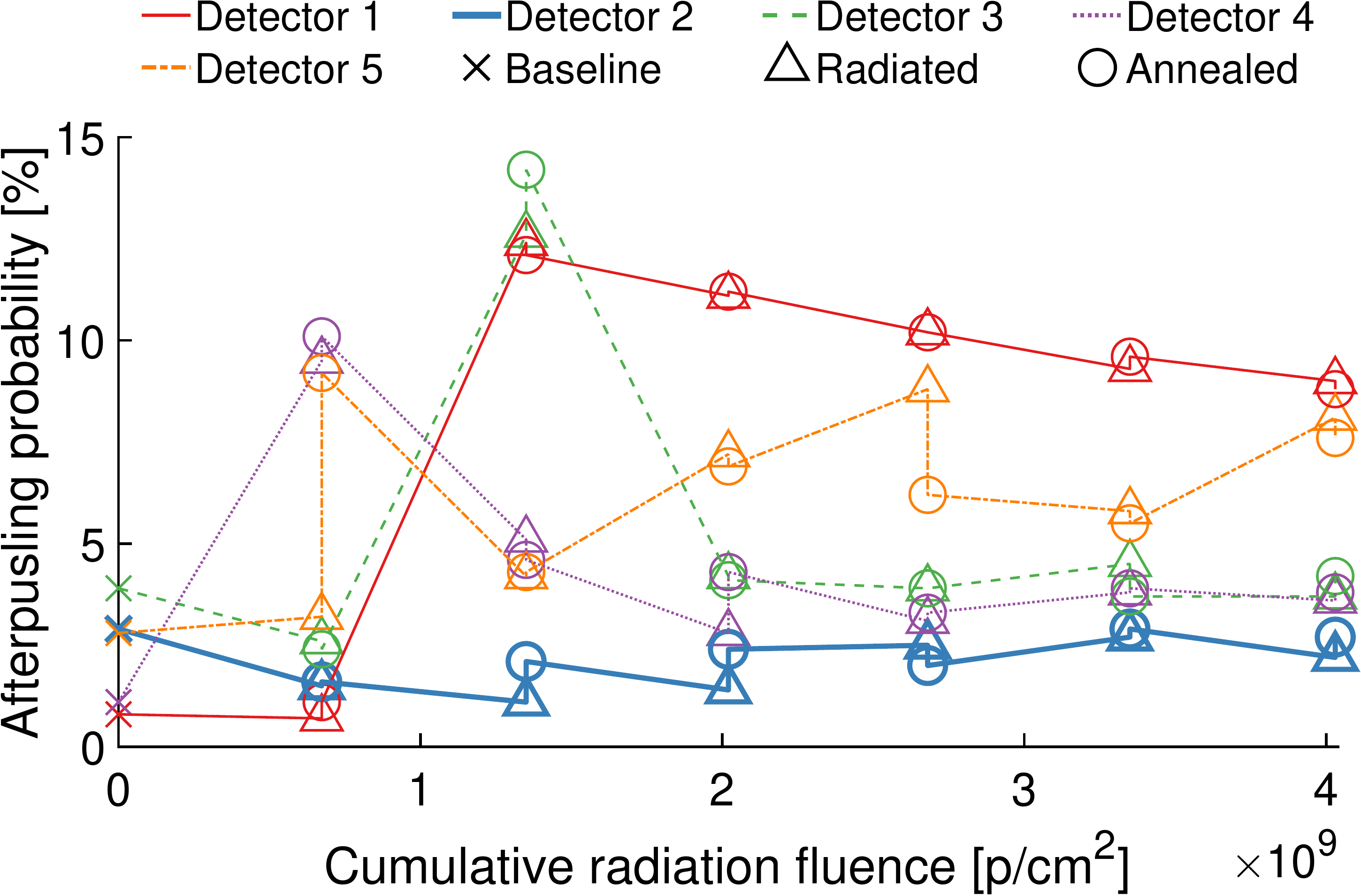}
  \caption[Afterpulsing probability for Radiation~1 plotted against radiation exposure]{Afterpulsing probability of DM1 detectors for cumulative radiation fluence. The afterpulsing probability at the end of the two-year-equivalent mark is generally higher than the baseline value, though a clear trend is not well exhibited by this data. No significant correlation with annealing is evident.}
  \label{fig.Afterpulsing_rad_1}
\end{figure}
 
\begin{figure}
  \includegraphics[width=\linewidth]{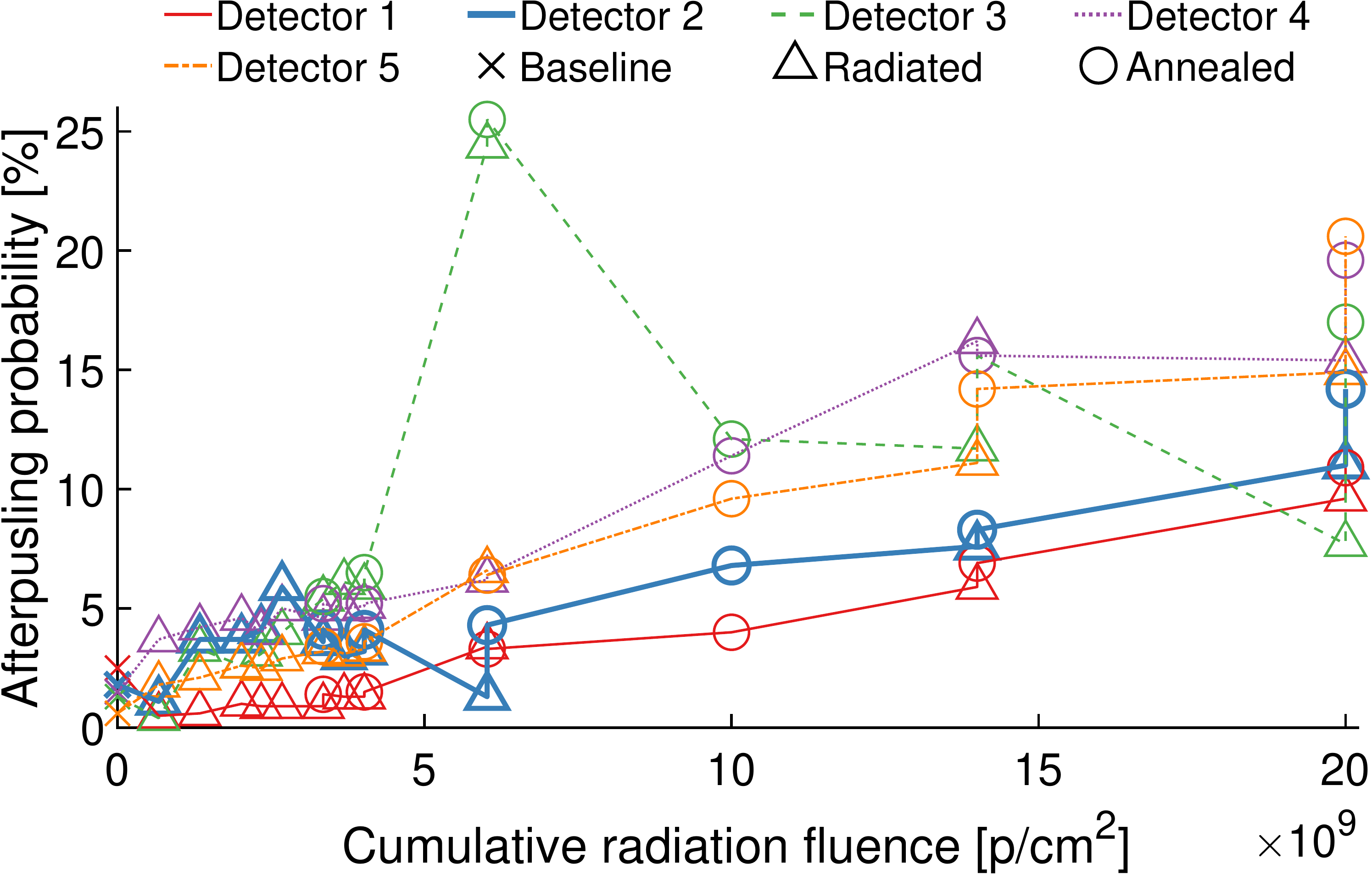}
 \caption[Afterpulsing probability for Radiation~2 plotted against radiation exposure]{Afterpulsing probability of DM2 detectors for cumulative radiation fluence. Larger incremental irradiations show a clear tendency towards higher afterpulsing probability. As with DM1, no significant correlation with thermal annealing is evident.}
  \label{fig.Afterpulsing_rad_2}
\end{figure}

Afterpulsing probabilities are calculated for both DMs by evaluating the distribution of time deltas between detection events and the corresponding laser reference pulse (\cref{sec:methodology}). Figures~\ref{fig.Afterpulsing_rad_1} and~\ref{fig.Afterpulsing_rad_2} show the afterpulsing probabilities evaluated corresponding to the characterizations performed after each incremental irradiation and annealing cycles for DM1 and DM2. Afterpulsing probability is seen to generally increase with each incremental irradiation, particularly for the higher fluences of DM2. Thermal annealing has no consistent effect on the afterpulsing probability. For DM1, afterpulsing probabilities are higher at \SI{4e9}{p/cm^2} cumulative fluence than the corresponding baseline (unirradiated) values for three of the five detectors. This effect is more pronounced in DM2, which is radiated up to \SI{2e10}{p/cm^2} cumulative fluence, where all the detectors registered higher afterpulsing probabilities compared to their baseline values.

Increased afterpulsing probability tends to increase the dark count rate, which could in part explain the increase of post-annealed dark count rates with higher cumulative fluence in \cref{fig.Post_annealed_dark_comparison}. Because of this tendency, high afterpulsing probabilities can undermine the usability of the detectors if the corresponding signal to noise ratio drops below a critical value. This may ultimately limit the lifetime of the satellite since it appears thermal annealing cannot mitigate the radiation-induced afterpulsing.

Afterpulsing arises from trapped charge carriers in defects of the detector substrate---proton radiation may increase the density of these defect sites. We note that the apparent increase in afterpulsing probability due to the radiation exposure does not appear to improve with annealing, unlike the dark count rate. To our knowledge, there is no established mechanism to explain this behaviour, and more research will be needed to identify this afterpulsing mechanism. While our detectors are passively quenched, the afterpulsing could be mitigated using methods established for telecom single-photon detectors, either using a gated operation (e.g., Ref.~\cite{he2017}), or free-running operation with active afterpulsing suppression (e.g., Ref.~\cite{tommaso2012}). Although we do not explore it here, these and related system parameters may be optimized for QKD key transmission.

\section{Conclusions}
\label{sec:conclusion}

The dark count rates of Si APD detectors increase substantially due to proton radiation experienced in low-Earth orbit. This can be mitigated by low (but well above cryogenic) temperature operation, and thermal annealing. By measuring the properties of detectors throughout multiple cumulative exposures to varying proton fluences, equivalent to different stages of a LEO life-cycle, and applying thermal annealing at some of these stages, we have established a profile representative of realistic operational scenarios of a quantum communication apparatus on a LEO space platform.

Our results show that even after repeated applications, annealing after irradiation is capable of achieving detector dark count rates suitable to perform QKD---below \SI{500}{cps} after radiation exposure of a nominal two-year mission lifetime, and potentially sufficiently low after even 10.5~years exposure, albeit with reduced performance in the latter case. Afterpulsing probability was shown to increase with cumulative fluence---however, thermal annealing had no significant effect on the afterpulsing probability. All other detector parameters were unaffected by both cumulative fluence and annealing.

Our observations further suggest that annealing conditionally---only when the dark count rate exceeds a threshold value (here \SI{2}{\kilo cps})---may achieve slightly lower post-annealed dark count rates over time in orbit than annealing at fixed periodic intervals. It should be noted that the optimal choice of threshold would depend significantly on the design and operational parameters of the QKD satellite platform that determine the expected channel losses and noise caused by background light. Further, a trade-off between lifetime and key rate can be made: applications interested in high performance over a short term could employ a lower threshold, thus performing annealing more frequently, compared to applications that can accept more moderate performance in exchange for a longer lifetime. Determining the appropriate balance would require additional quantitative analysis incorporating such details---the nonlinear response of the lifetime amount of generated key to the operating conditions may imply an interesting optimization problem. We leave this for future work.

We have demonstrated that thermal annealing is an effective strategy for mitigating radiation-induced dark counts, is feasible under the continuous radiation environment of a low-Earth orbit, and can be applied at appropriate times as a device maintenance procedure. This knowledge helps guide future developments, including the QEYSSat mission and beyond, towards cost-effective, high-performance, and long-lived quantum devices in space.

\acknowledgments
We thank Excelitas Technologies for discussions and for providing selected APD samples. We also thank NEPTEC Design Group for discussions, designing and building the DM. This work was supported by the Canadian Space Agency (STDP and FAST programs), Industry Canada, Canada Foundation for Innovation (CFI), Ontario Research Fund (ORF), Natural Sciences and Engineering Research Council (NSERC; programs Discovery and CryptoWorks21), and the Ontario Ministry of Research, Innovation and Science (MRIS).

\appendix
\section{Annealing time}\label{appendix.AnnealingTime}

\begin{figure}
	\includegraphics[width=\linewidth]{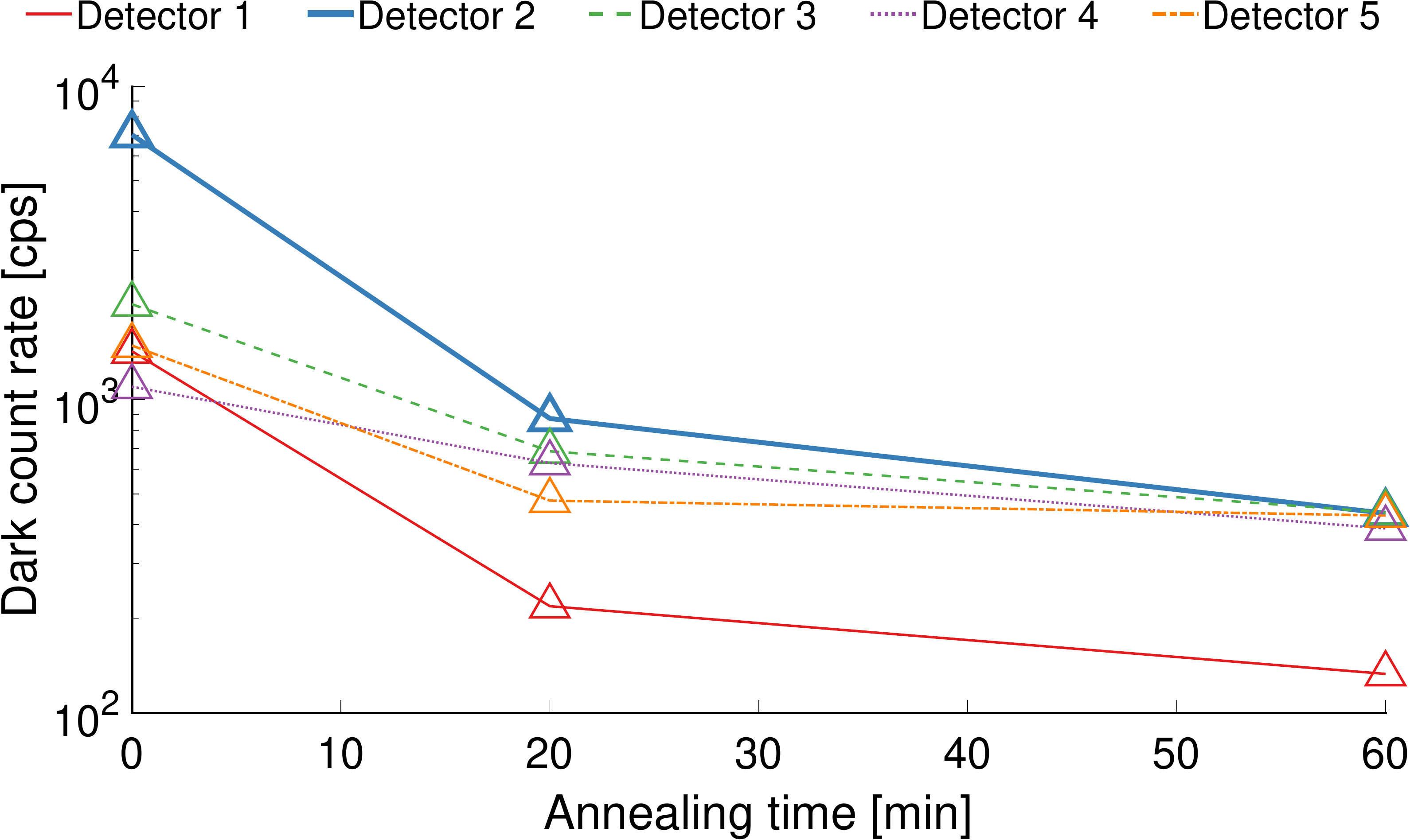}
	\caption[Annealing duration effect on Dark counts of Radiation~2 after \SI{6e9}{p/\cm^2}]{Post-annealing dark count rates dependence on annealing duration. The dark count rate was measured immediately following irradiation to the \SI{6e9}{p/\cm^2} cumulative fluence, again after \SI{20}{\minute} of annealing, and finally again after a further \SI{40}{\minute}, for \SI{60}{\minute} total annealing time (the cumulative time is shown here). The observed dark count reduction is considerably more significant in the first \SI{20}{\minute}.}
	\label{fig.Dark_rad_2_anneal_time}
\end{figure}

We chose a baseline annealing time of one hour, following results of previous work~\cite{Elena'sRadiationPaper}. However, for the sake of reproducibility, we also confirmed that our expectations were justified by taking an intermediate measurement at a shorter time while annealing DM2 at a total fluence of \SI{6e9}{p/\cm^2}. Fig.~\ref{fig.Dark_rad_2_anneal_time} shows the measured dark count rates corresponding to 20~minutes of annealing followed by an additional 40~minutes of annealing. The decrease in the dark count rate has significantly greater magnitude during the first \SI{20}{\minute} compared to the subsequent \SI{40}{\minute}, with the dark count rates beginning to stabilize by the 60~minute mark---indicating further annealing time would have little impact. This is consistent with the previous work~\cite{Elena'sRadiationPaper}.

\def\bibsection{\medskip\begin{center}\rule{0.5\columnwidth}{.8pt}\end{center}\medskip} 
\bibliography{bibtex_library}

\end{document}